\documentclass[a4paper,11pt]{article}
\pdfoutput=1 

\usepackage{jheppub} 

\usepackage{graphicx,amssymb,bm,latexsym,amsmath}
\usepackage{epstopdf}
\usepackage{caption}
\usepackage{subcaption}
\usepackage[section]{placeins}
\usepackage{float}
\usepackage{capt-of}
\usepackage{tabu}
\usepackage[toc,page]{appendix}

\newcommand{\nl}{\nonumber}
\newcommand{\mathsym}[1]{{}}
\newcommand{\unicode}[1]{{}}

\title{\boldmath Perturbation of Pulsating Strings}

 \author{Sorna Prava Barik,}
 \author{Kamal L. Panigrahi,}
 \author{Manoranjan Samal}
 \affiliation{Department of Physics,\\Indian Institute of Technology Kharagpur,\\
 	Kharagpur-721 302, India}

\emailAdd{sorna}
\emailAdd{panigrahi}
\emailAdd{manoranjan@phy.iitkgp.ernet.in}

\abstract{We discuss semiclassical quantization of circular pulsating strings in $ \text{AdS}_3 \times \text{S}^3 $ background with and without the Neveu-Schwarz- Neveu-Schwarz (NS-NS) flux. We find  the equations of motion corresponding to the quadratic action in bosonic sector in terms of  scalar quantities and invariants of the geometry. The general equations for studying physical perturbations along the string in an arbitrary curved spacetime are written down using covariant formalism. We  discuss the stability of these string configurations by studying the solutions of the linearized perturbed equations of motion.}

\begin{document}
\maketitle
\flushbottom
\section{Introduction}
The study of string  dynamics in curved space-time has always been an exciting area of research in string theory.  In the framework of AdS/CFT correspondence, the quantization of semiclassical string \cite{Gubser:2002tv} has been  extremely useful in establishing integrability structure \cite{Tseytlin:2010jv}  and dualities between two sides. The linearized perturbation of semiclassical string is helpful in matching the duality beyond the leading order. The main motivations behind studying perturbative  solutions are to investigate  the stability properties of the string solutions and to find the quantum string corrections to the expectation value of the Wilson loops. \par
 A non-covariant approach was introduced in \cite{deVega:1987veo,deVega:1988jh,deVega:1988ch}  to study the worldsheet fluctuation of string in curved space-time, especially in black hole or de-Sitter backgrounds  which led to finding the physical quantities like mass spectrum and scattering amplitudes. 
 On the other hand, a covariant approach in conformal gauge  was developed in \cite{Larsen:1993iva} and was used to find the perturbations of stationary strings embedded in Rindler, Schwarzschild and Reisner-Nordstr\"{\o}m spacetimes. It has been found that  the frequencies of first order as well as  second order fluctuations of the string  in (2+1)-dimensional black hole and black string backgrounds are real, which  shows  string  is stable in these backgrounds \cite{Larsen:1994ah}.   It was shown further that perturbations of circular strings in a power law expanding  universe grow much slower than the radius, implying the perturbation  is suppressed by the inflation of the universe \cite{Larsen:1994jt}. Similarly perturbation of the planetoid string in Ellis geometry and in (2+1)-dimensional BTZ black hole, shows the presence of world-sheet singularity at the edges \cite{Kar:1997zi}. All the above works were based on the perturbation of bosonic string action. \par
 
 The study of worldsheet fluctuation of the classical string in $\text{AdS}_5\times \text{S}^5$ background in Green-Schwarz formalism was developed in \cite{Drukker:2000ep,Drukker:2011za} which was an important step in extension of the AdS/CFT correspondence beyond the classical level. It was shown that in conformal and static gauge, the one loop correction to Green-Schwarz action can be expressed in terms of  differential operators, where the determinants of these operators give rise to well defined and finite partition function.
The quadratic quantum fluctuation of the rigidly rotating  homogeneous string solution\cite{Frolov:2003qc,Frolov:2003tu,Frolov:2004bh,Fuji:2005ry}  helps to find out the stability condition of the string dynamics. It has been shown that the pulsation enhances the stability of the multispin string solitons \cite{Khan:2005fc}.  In case of non-homogeneous string solution  \cite{Frolov:2002av, Beccaria:2010zn}  the quadratic fluctuation is expressed in terms of single-gap-Lam\'{e} operator. This approach has also been applied in $\text{AdS}$ backgrounds\cite{Forini:2012bb}. There are some instances where the two loop worldsheet fluctuations  have been carried out to match the subleading correction to  cusp anomalous dimension  of the strongly coupled gauge theory \cite{Roiban:2007jf,Kiosses:2014tua,Bianchi:2014ada}. The pohlmeyer reduction formalism has also been used to find the quantum fluctuation of the semiclassical strings in  $\text{AdS}_5\times \text{S}^5$  where it matches with the results of one loop computation \cite{Hoare:2009rq}. Some recent work in this line has been done in \cite{Forini:2015mca,Bhattacharya:2016ixc}.\par

The other interesting example of holographic dual pair is that of string propagation in $AdS_3 \times S^3 \times T^4$ background and the dual $N = (4,4)$ superconformal field theory. This duality has also been well explored from both sides and various semiclassical solutions in this background have been studied in the context of integrability \cite{Hoare:2013pma, Hoare:2013ida, Lloyd:2014bsa}. String theory in this background supported by NS-NS type flux can be described in terms of a $SL(2,R)$ Wess-Zumino-Witten model. It has been suggested recently that in $AdS_3\times S^3$ background supported by both NS-NS and RR fluxes $( H_3 = dB_2 ~{\rm and} ~F_3 = dC_2 )$, the string theory is integrable as well \cite{Cagnazzo:2012se, Wulff:2014kja}. In this context a class of semiclassical string solutions have been studied in some detail in \cite{Banerjee:2014gga, Hernandez:2014eta, Banerjee:2015bia}. The dynamics of such solutions in pure NS-NS flux case both in the large and small charge limit have been discussed. The fluctuations around rigidly rotating string in presence of NS-NS flux have been considered in \cite{Rashkov:2002zt,Dimov:2003bh,Larsen:2003ma}. It was shown that the presence of flux couple the fluctuation modes in a non-trivial way that indicates the changes in quantum correction to the energy spectrum. It would be interesting to understand the one loop corrections to the energy of pulsating strings at one loop level as well.

Motivated by recent surge of interest in studying semiclassical strings and their perturbations, in this paper we study the worldsheet perturbation of pulsating strings in  $\text{AdS}_3 \times  \text{S}^3$ in the presence of background NS-NS flux. The pulsating strings are one of simple string solutions whose gauge theory duals are known to exist and their anomalous
dimensions has been found out by \cite{Minahan:2002rc}. For very large
quantum numbers, the circular pulsating string expanding and
contracting on the S$^5$. These solutions are time-dependent as opposed to the usual rigidly rotating string solutions. They are
expected to be dual to highly excited states in terms of operators. For example the most general pulsating string in $S^5$ charged under the isometry group 
$SO(6)$ will have
a dual operator of the form ${\rm Tr}(
X^{J1}Y^{J2}Z^{J3})$, where $X, Y, {\rm and}~Z$, are the chiral
scalars and $J^i$'s are the R-charges of the SYM theory. Hence it will be interesting to know their fate
for small worldsheet fluctuations in the presence of NS-NS flux. 
The perturbation of spiky strings in flat and AdS space, in covariant formalism, has been discussed in \cite{Bhattacharya:2016ixc,Bhattacharya:2018unr}.

The rest of the paper is organized as follows. In section 2, we review the worldsheet perturbation formalism in the presence of background flux. In section 3 we construct  the analytical solution of the perturbation equation for pulsating string in $ R \times S^2 $ background in short string limit and comment on the stability of such solutions. Section 4 and 5 are devoted to study the physical perturbations of the pulsating strings in short string limit in $AdS_3$ background with and without three form fluxes. Finally, in section 6, we present our conclusion.
\section{Worldsheet perturbation of bosonic strings with NS-NS flux}
In this section, we recollect  the first order perturbation equations for bosonic string moving in an arbitrary curved spacetime using covariant approach. In this paper, we will follow the same approach as described in \cite{Garriga:1991ts,Guven:1993ew,Larsen:1993iva,Kar:1997zi,Viswanathan:1996yg,Larsen:2003ma}. 
We start with the Polyakov action for the bosonic string 
\begin{equation}\label{polyakov}
S_{P}=-\frac{1}{4 \pi \alpha^\prime}\int d \tau d \sigma\sqrt{-h} h^{\alpha \beta }G_{\alpha \beta},
\end{equation}
where $h_{\alpha \beta}$ is the internal metric and $G_{\alpha \beta}$ is the induced metric on string worldsheet:
\begin{equation}
G_{\alpha \beta}= g_{\mu \nu} \frac{\partial X^\mu}{\partial \sigma^ \alpha} \frac{\partial X^\nu}{\partial \sigma^ \beta}.
\end{equation}
Here $\sigma^\alpha(\alpha=0,1)$ are the worldsheet coordinates whereas $X^\mu(\mu=0,1,2,\cdots,n)$ are the spacetime coordinates. The equations of motion corresponding to the action (\ref{polyakov}), take the form
\begin{equation}\label{eom}
\ddot{X}^\mu -X^{\prime \prime^\mu}+\Gamma^\mu_{~\rho \sigma}\left(  \dot{X}^\rho \dot{X}^\sigma-X^{\prime^\rho } X^{\prime \sigma}\right)=0,
\end{equation}
where dot and prime denote the derivatives with respect to $\tau$ and $\sigma$ respectively.
This set of equations of motion is supplemented by conformal gauge constraint,
\begin{equation}\label{constriant}
G_{\alpha \beta}- \frac{1}{2}h_{\alpha\beta}G^\gamma_{~\gamma}=0.
\end{equation}
 In order to get the first order perturbation equation  we vary (\ref{eom}). It can also be achieved by taking second order variation of the action (\ref{polyakov}). Since the normal and  tangential vectors are the basis vectors of the spacetime coordinates, we can decompose the variation of the spacetime coordinates  as sum of projections on them,
\begin{equation}
\delta X^ \mu = \varphi^i N^\mu_i+ \psi^ \alpha X^\mu{,_\alpha}
\end{equation}
Here ``${X^{\mu}}_{,\alpha}$'' is the
derivative of $X^{\mu}$ with respect to the worldsheet coordinates $\alpha$. Further, $N^\mu_i $and $ X^\mu,_\alpha $ are the normal and tangential vectors respectively. The $\varphi^i$ are scalars where $i$ is a number index  and $\psi^\alpha$ are just reparametrizations.\\
The normal vectors satisfy the following orthogonality condition:
\begin{equation}\label{conditions}
g_{\mu \nu} N^ \mu_i N^\nu_j= \delta_{ij}  ~~~~~~~~ \text{and}  ~~~~~~~ g_{\mu\nu}N^\mu_i \partial_\alpha X^\nu=0.
\end{equation}
The action remains invariant for the tangential projection $(\psi^ \alpha X^\mu{,_\alpha})$ due to worldsheet diffeomorphism. Hence only projection on normal vectors contribute to the physical perturbation.
Let us introduce the extrinsic curvature tensor($K^i_{~\alpha \beta}$) and the normal fundamental form($\mu_{i j}^{~~\alpha}$)
\begin{eqnarray}
K_{i,\alpha \beta}=  g_{\mu \nu} N_i^\mu X^\rho_{,\alpha} \triangledown_\rho X^ \nu_{,\beta}, ~~~~~
\mu_{i j, \alpha}= g_{\mu \nu} N^\mu_{i} \partial_\alpha X^ \rho\triangledown_\rho N^ \nu_{j},
\end{eqnarray}
where $\triangledown_\rho$ is the covariant derivative with respect to spacetime coordinate.
The extrinsic curvature tensor are symmetric in world sheet coordinates i.e. $K_{i,\alpha \beta}=K_{i,\beta \alpha}$ whereas the normal fundamental forms are antisymmetric in number indices i.e. $\mu_{i j}^{~~\alpha}=-\mu_{ji}^{~~\alpha}$.
The unperturbed equation of motion in (\ref{eom}) can also be expressed in terms of extrinsic curvature tensor as:
\begin{equation}
h^{\alpha \beta} K^i_{~\alpha \beta}=0.
\end{equation}
Finally we write the first order perturbation equation in terms of $\varphi^j$ as
\begin{eqnarray}\label{1stperteqn}
\Box \varphi_i +2 \mu_{i j}^{~~\alpha} \varphi^j_{, \alpha} +(\triangledown_\alpha \mu_{i j}^{~~\alpha}) \varphi^j- \mu_{i l}^{~~\alpha}  \mu_{j ~\alpha}^ {~l} \varphi^j\nl \\ +\frac{2}{G^c_{~c}}K^{~\alpha \beta}_i K_{j, \alpha \beta}\varphi^j-h^{\alpha \beta} R_{\mu \rho \sigma\nu } N^\rho_i N^\sigma_j X^\mu_{,\alpha} X^ \nu _{, \beta}\varphi^j=0,
\end{eqnarray}
 where $\triangledown_\alpha$ is the covariant derivative with respect to internal metric. One can also write the perturbation equation in a compact form as
\begin{equation}
\left(\delta^{kl} h^{\alpha \beta} D_{ik \alpha}D_{lj \beta}+\frac{2}{G^c_{~c}}K^{~\alpha \beta}_i K_{j, \alpha \beta}-h^{\alpha \beta} R_{\mu \rho \sigma\nu } N^\rho_i N^\sigma_j X^\mu_{,\alpha} X^ \nu _{, \beta}\right)\varphi^j=0,
\end{equation} 
where $D_{ij \alpha}=\delta_{ij} D_{\alpha}+\mu_{i j \alpha}$. \par In general, the above equations give  a set of  coupled second order partial differential  wave  equations for scalar quantities.  These coupled  wave equations are manifestly covariant under worldsheet diffeomorphisms. \par

Now we wish to find the perturbation equation for the bosonic string embedded in an arbitrary background in the presence of nonvanishing NS-NS B-field. The relevant sigma-model action in conformal gauge
\begin{equation}\label{nambu}
S=-\frac{1}{4 \pi \alpha^\prime} \int d \tau d \sigma \left( \sqrt{-h}h^{a b}g_{\mu \nu} \partial_a X^\mu\partial_b X^\nu-\epsilon^{\alpha \beta} B_{\mu \nu}X^\mu ,_{\alpha} X^\nu,_{ \beta} \right).
\end{equation}
 Due to the presence of  non-vanishing B-field, the extrinsic curvature tensor is modified to,
\begin{equation}
\hat{K}_{\alpha \beta}=  \hat{\triangledown}_\rho (X^ \mu_{,\alpha}) X^\rho_{,\beta} N_\mu= \left(X^\mu_{,\alpha \beta}+\hat{\Gamma}^\mu_{\rho \sigma} X^\rho_{,\alpha} X^\sigma_\beta\right)N^\nu g_{\mu \nu},
\end{equation}
  where $\hat{\Gamma}^\mu_{\rho \sigma}$ is the modified Christoffel symbol defined as
\begin{equation}
\hat{\Gamma}^\mu_{\rho \sigma}=\Gamma^\mu_{\rho \sigma}-\frac{1}{2}g^{\mu \nu}\left(\partial_\nu B_{\rho \sigma}+\partial_\rho B_{\sigma \nu}+\partial_\sigma B_{\nu \rho}\right).
\end{equation}
The equation of motion can also be written in terms modified extrinsic curvature tensor $\hat{K}_{\alpha \beta}$,
\begin{equation}
\left(h^{\alpha \beta}- \frac{\epsilon^{\alpha \beta}}{\sqrt{-h}}\right ) \hat{K}_{\alpha \beta }=0.
\end{equation}
The first order perturbation equation turns out to be

\begin{equation}\label{2ndperteq}
\left(\delta^{kl} h^{\alpha \beta} D_{ik \alpha}D_{lj \beta}+\frac{2}{G^c_{~c}}\hat{K}^i_{~\alpha \beta}\hat{K}^{~\alpha \beta}_j - \left(h^{\alpha \beta}-\dfrac{\epsilon^{\alpha \beta}}{\sqrt{-h}} \right) \hat{R}_{\sigma \nu \mu \rho} X^\mu_{~,\alpha} X^\nu_{~,\beta} N^ {i\rho} N^ \sigma_j\right)\varphi^j=0,
\end{equation}
where $D_{ij \alpha}=\delta_{ij} D_{\alpha}+\hat{\mu}_{i j \alpha}$.
\section{Perturbation of pulsating string in $R \times S^2$}
In this section we study the worldsheet perturbation of the pulsating string in $R\times S^2 \subset AdS_5 \times S^5$. We start with $R \times S^2$ metric
\begin{equation}
ds^2=-dt^2+d\theta^2+\sin^2 \theta d \phi^2.
\end{equation}
The string exhibits pulsating motion in one of the angular directions whereas it winds over the other direction, so the natural choice for such string embedding can be written as
\begin{equation}
t=k \tau, ~~~ \theta= \theta(\tau),~~~ \phi= m \sigma,
\end{equation}
where $m$ is the winding number.
The corresponding Polyakov action becomes,
\begin{equation}
S=-\frac{1}{4\pi \alpha\prime}\int d \tau d\sigma [\dot{t}^2-\dot{\theta}^2+ m^2 \sin^2 \theta ].
\end{equation}
The non-trivial equation of motion and conserved energy are given by
\begin{equation}\label{rs2eom}
\ddot{\theta}+m^2 \sin \theta \cos \theta =0,~~~~~~~~~~\mathcal{E}=-P_t=\frac{\partial  \mathcal{L}}{\partial \dot{t}}=k.
\end{equation}
The conformal gauge constraint can be written as
\begin{equation}\label{rs2consraint}
\mathcal{E}^2=\dot{\theta}^2+m^2 \sin^2 \theta.
\end{equation}
The solution of $\theta$ for $\theta(0)=0$ can be expressed in terms of Jacobi elliptic function
\begin{equation}\label{rs2solution}
\sin \theta= \frac{\mathcal{E}}{m}sn\left(m \tau | \frac{\mathcal{E}^2}{m^2}\right).
\end{equation}
The induced metric on the worldsheet is given by
\begin{equation}
ds^2=m^2 \sin^2\theta (-d\tau^2 +d\sigma^2).
\end{equation}
The Ricci scalar of the  worldsheet metric turns out to be:
\begin{equation}
^2 R= -2 \frac{(\mathcal{E}^2 -m^2 \sin^4 \theta)}{m^2 \sin^4 \theta}.
\end{equation}
We  note that the worldsheet has singularity at $\tau=\frac{2n}{m} \mathbb{K}[\frac{\mathcal{E}^2}{m^2}]$, where n is integer and $\mathbb{K}[\frac{\mathcal{E}^2}{m^2}]$ is the complete Elliptic integral of first kind. . \par
Now let us  proceed to find the first order perturbation equation as described in the earlier section. The tangent vectors are
\begin{equation}
\dot{X}^\mu =(-\mathcal{E},~ \dot{\theta},~ 0),~~~~~X'^{\mu}= (0,~0,~m).
\end{equation}
Using  the orthogonality condition as described in (\ref{conditions}) we get the normal vector
\begin{equation}\label{rs2normalvec}
N^\mu=\left(\frac{\sqrt{\mathcal{E}^2-m^2 \sin^2 \theta}}{m \sin \theta},~~\frac{\mathcal{E}}{m \sin \theta},~~0\right).
\end{equation}
The components of extrinsic curvature tensor are given by
\begin{eqnarray}
K_{\tau \tau}=-m \mathcal{E} \cos \theta,~~~~~ K_{\tau \sigma}=0, ~~~~~
K_{\sigma \tau}=0,~~~~~~ K_{\sigma \sigma}=-m \mathcal{E} \cos \theta.
\end{eqnarray}
Here all the normal fundamental forms vanish.
Using (\ref{1stperteqn}) we get the first order  perturbation equation
\begin{equation}
-\ddot{\varphi}(\tau,\sigma)+\varphi^{\prime \prime}(\tau ,\sigma)+ (2 \mathcal{E}^2\cot^2 \theta +\mathcal{E}^2) \varphi(\tau ,\sigma)=0.
\end{equation}
Then using  the Fourier expansion method, i.e. substituting $\varphi(\tau, \sigma)= \sum_n\alpha_n(\tau) e^{i n \sigma}$ in the previous equation, we get
\begin{equation}\label{rs2pert}
\ddot{\alpha}_n+\left(n^2- \frac{2 \mathcal{E}^2}{\sin^2 \theta} +\mathcal{E}^2 \right)\alpha_n=0.
\end{equation}
In case of short string limit, i.e. $\mathcal{E}<<m$, we drop quadratic and higher order term of $\frac{\varepsilon}{m}$. So the solution of $\theta$ becomes
\begin{equation}
\sin \theta=\frac{\mathcal{E}}{m}\sin(m \tau).
\end{equation}
Therefore the perturbation equation turns out to be
\begin{equation}{\label{rs2pert with short string limit}}
\ddot{\alpha}_n+\left(n^2- \frac{2 m^2}{\sin^2 m \tau} +\mathcal{E}^2 \right)\alpha_n=0.
\end{equation}
which is a  special case of well-known Poschl-Teller equation and it's general solution can be written in terms of hypergeometric functions
\begin{equation}
\alpha_n=C_1  P_1(\tau)+C_2 P_2(\tau),
\end{equation}
where $$P_1(\tau)=\sin^2 m \tau{}~_2F_1\left(1+\frac{\sqrt{n^2 +\mathcal{E}^2}}{2 m},1-\frac{\sqrt{n^2 +\mathcal{E}~^2}}{2 m},\frac{5}{2};sin^2 m \tau\right), $$ 
$$~~ P_2(\tau)=\frac{1}{\sin m \tau}~ {}_2F_1\left(\frac{\sqrt{n^2 +\mathcal{E}^2}-m}{2 m},\frac{-\sqrt{n^2 +\mathcal{E}^2}-m}{2 m},-\frac{1}{2};sin^2 m \tau\right) $$
$\text{and } C_1 , C_2 $ are constants. The scalar function $\varphi(\tau, \sigma)$ in the perturbation equation takes the form
\begin{equation}
 \varphi(\tau, \sigma)=\sum_n\varepsilon_0(C_1  P_1(\tau)+C_2  P_2(\tau)) e^{i n \sigma}.
\end{equation}
It should be mentioned  that  $\varepsilon_0$  $(\varepsilon_0<<1)$ is a constant and it can be related as the  amplitude of the perturbation. Here we can note that for every allowed value of $n$, the first solution is a oscillatory while the second one is periodically diverging. Observing the  [fig.1], one can comment that  when  $n$ increases amplitude of  perturbation decreases for $P_1$. We can also see that when the string becomes point-like,  $P_1$ vanishes while $P_2$ blows up which gives rise to instability. The  physical perturbations  for short string solution can be written as
\begin{eqnarray}
\delta t&=& N^t \varphi(\tau, \sigma) =\varepsilon_0\sum_n   \sin m\tau \cos m\tau \cos n\sigma~~ {}_2F_1\left(1+\frac{\sqrt{n^2 +\mathcal{E}^2}}{2 m},1-\frac{\sqrt{n^2 +\mathcal{E}^2}}{2 m},\frac{5}{2};sin^2 m \nonumber \tau\right), \\ \nonumber
\delta \rho &=& N^\rho \varphi(\tau, \sigma)=\varepsilon_0 \sum_n  \sin m\tau \cos n\sigma~~ {}_2F_1\left(1+\frac{\sqrt{n^2 +\mathcal{E}^2}}{2 m},1-\frac{\sqrt{n^2 +\mathcal{E}^2}}{2 m},\frac{5}{2};sin^2 m \nonumber \tau\right), \\ \nonumber
\delta \theta &=&N^\theta \varphi(\tau, \sigma)= 0.
\end{eqnarray}
Here we take the real part i.e $\cos n\sigma$ of $\sigma$ dependent term and the oscillatory solution of the $\tau$ dependent term. 
\begin{figure}[H]
	\centering
	\begin{subfigure}[t]{0.5\textwidth}
		\centering
		\includegraphics[width=1\textwidth]{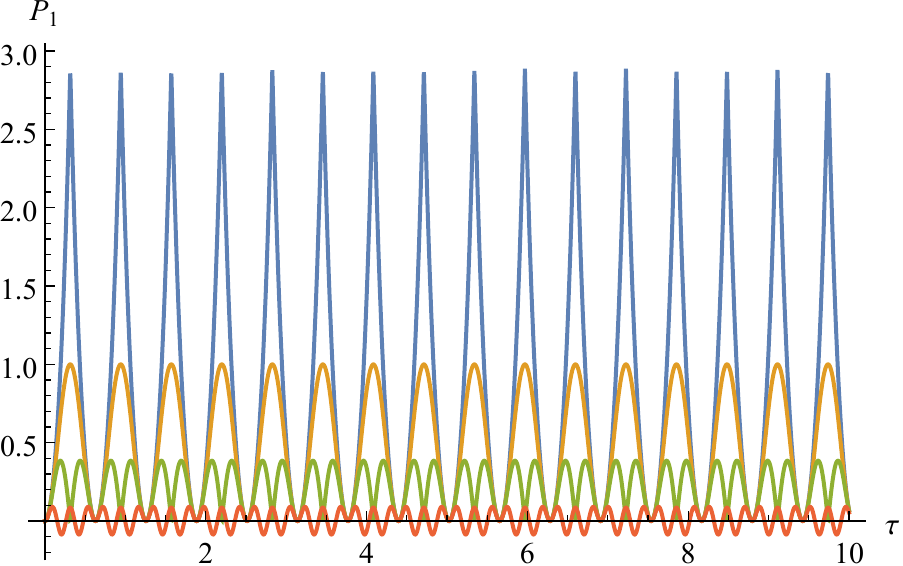}
		\caption{}
	\end{subfigure}%
	~ 
	\begin{subfigure}[t]{0.6\textwidth}
		\centering
		\includegraphics[width=1\textwidth]{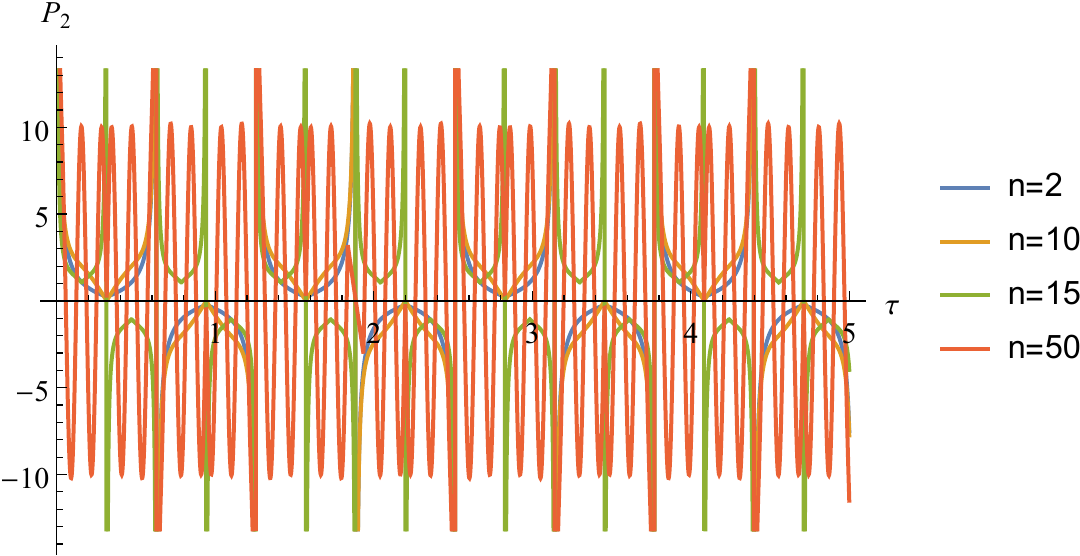}
		\caption{}
	\end{subfigure} 
	
	\caption{ Perturbations for $m=5$, $\mathcal{E}=0.03$    }
\end{figure}

Now let us  discuss perturbation  equation for $\varphi(\tau,\sigma)$ without any approximation. Substituting $sn^2(m\tau,\frac{\mathcal{E}^2}{m^2})=s$ in the equation (\ref{rs2pert}), the equation takes the form
\begin{equation}\label{heun1}
\frac{\partial^2 \alpha_n}{\partial s ^2}+\frac{1}{2}\left(\frac{1}{s}+\frac{1}{s-1}+\frac{1}{s-\frac{m^2}{\mathcal{E}^2}}\right)\frac{\partial\alpha_n}{\partial s}+\frac{1}{s(s-1)(s-\frac{m^2}{\mathcal{E}^2})}\left(\frac{n^2 +\mathcal{E}^2}{4\mathcal{E}^2}-\frac{m^2}{2\mathcal{E}^2 s}\right)\alpha_n=0.
\end{equation}
Since the differential equation have four regular singular points i.e $ 0,1,\frac{m^2}{\mathcal{E}^2} ~\text{and}~  \infty $, it can be converted into Heun's equation by proper transformation. 
Now putting  $\alpha_n=u s$ in equation (\ref{heun1}), we have
\begin{equation}
\frac{\partial^2 u}{\partial s ^2}+\frac{1}{2}\left(\frac{5}{s}+\frac{1}{s-1}+\frac{1}{s-\frac{m^2}{\mathcal{E}^2}}\right)\frac{\partial u}{\partial s}+\frac{1}{s(s-1)(s-\frac{m^2}{\mathcal{E}^2})}\left(\frac{3}{2}s+\frac{n^2 -4m^2}{4\mathcal{E}^2}-\frac{3}{4}\right)u=0.
\end{equation}
which is exactly the canonical form of Heun's general equation(see the Appendix) and the solution to this equation  can be written in terms of Heun's function. In general the Heun's function can be expanded in terms of hypergeometric functions, but we shall not go into details here. So finally the solution of the perturbation equation becomes
\begin{equation}
\alpha_n(\tau)=\varepsilon_0~ sn^2(m\tau,\frac{\mathcal{E}^2}{m^2})~~ H\left(\frac{m^2}{\mathcal{E}^2},\frac{3}{4}-\frac{n^2-4m^2}{4 \mathcal{E}^2};\frac{3}{2},1,\frac{5}{2},\frac{1}{2};sn^2(m\tau,\frac{\mathcal{E}^2}{m^2})\right).
\end{equation} 
As before, we have included $\varepsilon_0$ here which is the amplitude 
of perturbation.
\section{Perturbation of pulsating string in $AdS_3$}
In this section we consider the pulsating string in $AdS_3$ background and then we find the equation of motion of the worldsheet perturbation. The relevant metric is given by
\begin{equation}
ds^2= -\cosh^2 \rho dt^2+d\rho^2+\sinh^2 \rho d\theta^2.
\end{equation}
The ansatz for the pulsating string:
\begin{equation}
t=t( \tau), ~~~~~~~~~\rho= \rho(\tau),~~~~~~~~\theta= m \sigma .
\end{equation}
The string is pulsating along the radial direction of $AdS$ space  whereas it spreads over the $\theta$ direction with winding number $m$.
We can find the equation of motion for $\rho$,
\begin{equation}
\ddot{\rho}+\cosh\rho \sinh \rho(\dot{t}^2+ m^2)=0.
\end{equation}
The conserved $AdS$ energy is given by,
\begin{equation}
\mathcal{E}=-\frac{\partial \mathcal{L}}{\partial \dot{t}}= \dot{t} \cosh^2 \rho.
\end{equation}
Now we can write Virasoro constraint equation in terms of $AdS$ energy
\begin{equation}\label{ads3constraint}
\dot{\rho}^2-\frac{\mathcal{E}^2}{\cosh^2 \rho}+m^2\sinh^2 \rho =0.
\end{equation}
To find the explicit form of $\rho (\tau)$ coordinate, we need  to  solve the Virasoro constraint equation.
 So for $\rho(0)=0$,  the solution for  $\rho$ becomes
\begin{equation}\label{ads3solution}
\sinh \rho = \sqrt{\frac{-R_+ R_-}{R_+-R_-}}sd\left(m \sqrt{R_+- R_-} \tau  |  \frac{R_+}{R_+-R_-} \right),
\end{equation}
where
\begin{eqnarray}
R_\pm=\frac{-m\pm\sqrt{m^2+4\mathcal{E}^2}}{2m}.
\end{eqnarray}
We can note that the solution is periodic since the value of $\frac{R_+}{R_+-R_-}\textless1$.
The value of $\rho$ starts from zero and it reaches maximum value $\rho_{max}=\sinh^{-1}\sqrt{R_+}$ and again contracts to zero and so on. The periodic nature of the solution shows that the string is pulsating in $\rho$ direction. The induced metric is given by,
\begin{equation}
ds^2= m^2 \sinh^2 \rho [-d\tau^2+d\sigma^2].
\end{equation}
The curvature scalar of the worldsheet metric turns out to be:
\begin{equation}
^2 R=-2 \frac{(\mathcal{E}^2 + m^2 \sinh^4 \rho)}{m^2 \sinh^4 \rho}.
\end{equation}
The curvature scalar shows that there is singularity at the center of AdS which corresponds to $\frac{2n}{m\sqrt{R_+-R_-}}\textbf{K}\left(  \frac{R_+}{R_+-R_-}\right) $ value of $\tau$. When the string passes through the center of AdS it
becomes pointlike and the world-sheet has a curvature singularity.
The tangent vectors are given by 
\begin{equation}
\dot{X}^\mu= \left( \frac{\mathcal{E}}{\cosh^2 \rho},~~~~\sqrt{\frac{\mathcal{E}^2}{\cosh^2 \rho}-m^2 \sinh^2 \rho},~~~0\right),~~~X'^\mu=(0,~~0,~~m).
\end{equation}
The normal vector satisfying the orthogonality relation (\ref{conditions}) is
\begin{equation}
N^ \mu = \left( \frac{\sqrt{\mathcal{E}^2 -m^2 \sinh^2 \rho \cosh^2 \rho}}{m \sinh \rho  \cosh^2 \rho}, ~~ \frac{\mathcal{E}}{m  \sinh\rho \cosh \rho},~~0\right).
\end{equation}
The components of extrinsic curvature tensor are,
\begin{eqnarray}
K_{\tau \tau} = -m \mathcal{E}, ~~ ~~ K _{\tau \sigma}=0,~~~~~
K_{\sigma \tau}=0, ~~~~~ K_{\sigma \sigma}= -m \mathcal{E}.
\end{eqnarray}
Now using (\ref{2ndperteq}), we can write first order perturbation equation as
\begin{equation}\label{ads3pert}
-\ddot{\varphi}(\tau,\sigma)+\varphi^{\prime \prime}(\tau, \sigma) + \left(\frac{2 \mathcal{E}^2}{\sinh^2 \rho}-2 m^2 \sinh^2 \rho \right)\varphi(\tau,  \sigma)=0.
\end{equation}
Using Fourier expansion method i.e.  $\varphi(\tau, \sigma)=\sum_n \alpha_n(\tau) e^{i n \sigma}$ and substituting the value of $\sinh \rho $  from (\ref{ads3solution}), the perturbation equation takes the form
	\begin{equation}
\ddot{\alpha}_n(\tau)+\left(n^2-\frac{2\mathcal{E}^2}{a^2 sd^2(b \tau,c)}+2m^2a^2 sd^2(b \tau,c) \right)\alpha_n(\tau)=0,
	\end{equation}
where $$a=\sqrt{\frac{-R_+ R_-}{R_+-R_-}}, ~~b=m \sqrt{R_+-R_-}   ~~\text{and} ~~
c= \frac{R_+}{R_+-R_-}.$$
The above equation is a generalized Lam\'{e} equation or Heun equation with four regular singularities whose general  solution can be written in terms of infinite sum of hypergeometric functions. But here we will not discuss details about it,  we will consider the perturbation only for special case i.e. for  short string solution. \par
For short string limit $\varepsilon<<m$, the solution for $\rho$ reduces to
 \begin{equation}
 \sinh \rho \approx \frac{\mathcal{E}}{m} \sin m \tau.
 \end{equation}
 Taking the above approximation, the perturbation equation for short string limit turns out be
  \begin{equation}
  \ddot{\alpha}_n(\tau)+\left(n^2-\frac{2m^2}{\sin^2 m \tau} \right)\alpha_n(\tau)=0,
 \end{equation} 
 The general solution of the above equation can be written in terms of linear combination of two hypergeometric functions
 \begin{equation}
\alpha_n=C_1 \sin^2 m \tau ~P_1(\tau)+C_2 \frac{1}{\sin m \tau} ~P_2(\tau)
 \end{equation}
 where $$P_1(\tau)={}_2F_1\left(1+\frac{n}{2 m},1-\frac{n}{2 m},\frac{5}{2};sin^2 m \tau\right)~~~\text{and}~~~P_2(\tau)={}_2F_1\left(\frac{n-m}{2 m},\frac{-n-m}{2 m},-\frac{1}{2};sin^2 m \tau\right).$$
 Our general solution is the combination two independent solutions for any value  of $n$, out of which one is oscillatory and other one is divergent solution.
 The scalar function $\varphi(\tau, \sigma)$ in the perturbation equation takes the form
 \begin{equation}
 \varphi(\tau, \sigma)=\varepsilon_0 \sum_n\left[C_1 \sin^2 m \tau ~P_1(\tau)+C_2 \frac{1}{\sin m \tau} ~P_2(\tau)\right]e^{i n \sigma}.
 \end{equation}
 As described in $R\times S^2$ case, here also the diverging term in the perturbation  gives instabilities to the solution.  Finally we write the physical perturbation to the coordinates of $AdS_3$ by using oscillatory solution of the perturbation equation
\begin{eqnarray}
\delta t&=& N^t \varphi(\tau, \sigma) =\varepsilon_0  \sum_n \sin m\tau \cos m\tau \cos n\sigma~~ {}_2F_1\left(1+\frac{n}{2 m},1-\frac{n}{2 m},\frac{5}{2};sin^2 m \nonumber \tau\right), \\ \nonumber
\delta \rho &=& N^\rho \varphi(\tau, \sigma)=\varepsilon_0 \sum_n  \sin m\tau \cos n\sigma~~ {}_2F_1\left(1+\frac{n}{2 m},1-\frac{n}{2 m},\frac{5}{2};sin^2 m \nonumber \tau\right), \\ \nonumber
\delta \theta &=&N^\theta \varphi(\tau, \sigma)= 0.
\end{eqnarray}

\section{ Pulsating string in $AdS_3$ with NS-NS B-field}
In this section we include the nonvanishing NS-NS B-field for pulsating string in $AdS_3$ background. The relevant metric and NS-NS B-field are given by
\begin{eqnarray}
ds^2=-\cosh^2\rho dt^2 +d\rho^2 +\sinh^2\rho d\phi^2 \nonumber, \\
B_{t\phi}=q \sinh^2\rho,
\end{eqnarray}
where the parameter q takes value from 0 to 1. For 0 and 1 value of q the background is supported by pure RR flux and pure NS-NS flux respectively. Now taking the ansatz, $t=t(\tau), ~ \rho=\rho(\tau), ~\phi=m \sigma,$ the  corresponding  action becomes
\begin{eqnarray}
I=\frac{1}{4\pi\alpha\prime}\int d\tau d\sigma [-\cosh^2\rho\dot{t}^2+\dot{\rho}^2- m^2\sinh^2\rho+2q m \sinh^2\rho \dot{t} ].
\end{eqnarray}
Equations of motion for t and $\rho$ are given by
\begin{eqnarray} \label{Beom}
\dot{t} \cosh^2\rho-\mathcal{E}-qm\sinh^2\rho=0, \nonumber \\
\ddot{\rho}+ \sinh\rho \cosh\rho (\dot{t}^2+m^2-2qm\dot{t})=0,
\end{eqnarray}
where $\mathcal{E}$ is the energy of AdS. The Virasoro constraint equation reads,
\begin{eqnarray}\label{VC}
-\cosh^2\rho \dot{t}^2+\dot{\rho}^2+m^2\sinh^2\rho=0.
\end{eqnarray}
From equations (\ref{Beom}) and (\ref{VC}), we get the expression for $\rho$
\begin{equation}
\dot{\rho}^2=\dfrac{(\mathcal{E}+qm\sinh^2\rho)^2}{\cosh^2\rho}-m^2 \sinh^2\rho
\end{equation}
For $\rho(0)=0$ the above equation can be solved in terms Jacobi elliptic function,
\begin{eqnarray}\label{ads3solutionb}
\sinh \rho = \sqrt{\frac{-R_+ R_-}{R_+-R_-}}sd\left(m \sqrt{1-q^2} \sqrt{R_+- R_-} \tau  |  \frac{R_+}{R_+-R_-} \right),
\end{eqnarray}
where
\begin{eqnarray}
R_\pm=\frac{2 q \mathcal{E}-m\pm\sqrt{m^2 -4m q \mathcal{E}+4 \mathcal{E}^2}}{2m(1-q^2)}.
\end{eqnarray}
The induced metric and the curvature scalar are given by
\begin{eqnarray}
ds^2= m^2 \sinh^2\rho  \eta_{ab},~~~~^2 R=-2 \frac{(\mathcal{E}^2 +(1-q^2) m^2 \sinh^4 \rho)}{m^2 \sinh^4 \rho}.
\end{eqnarray}
The tangent and normal vectors are given by
\begin{eqnarray}
\dot{X}^\mu &=& \left( \frac{\mathcal{E}+q m \sinh^2 \rho}{\cosh^2 \rho},~~\sqrt{\dfrac{(\mathcal{E}+qm\sinh^2\rho)^2}{\cosh^2\rho}-m^2 \sinh^2 \rho},~~0\right),~~X'^\mu=(0,~~0,~~m), \nonumber \\
N^\mu &=&\left(\dfrac{\sqrt{(\mathcal{E}+qm\sinh^2\rho)^2-m^2 \sinh^2{\rho} \cosh^2\rho}}{m \sinh\rho \cosh^2\rho},\quad\dfrac{\mathcal{E}+qm \sinh^2\rho}{m \sinh\rho \cosh\rho},\quad 0 \right).
\end{eqnarray}
The components of extrinsic curvature tensor are given by
\begin{eqnarray}
K_{\tau\tau} = -m (\mathcal{E}-mq\sinh^2\rho), K_{\sigma\sigma} = -m (\mathcal{E}+mq\sinh^2\rho),  K_{\tau\sigma} = m^2q\sinh^2\rho=- K_{\sigma\tau}. \nonumber \\
\end{eqnarray}
Finally we write down the perturbation equation keeping up to 2nd order in $\mathcal{E}$ and first order in $q$
\begin{eqnarray}
\Box\varphi(\tau,  \sigma)+2\dfrac{(\mathcal{E}^2-m^2 \sinh^4\rho)}{\sinh^2\rho}\varphi(\tau,  \sigma)-2qm\mathcal{E}\varphi(\tau,  \sigma)=0.
\end{eqnarray}
Proceeding in the same way as the previous section and substituting    $\varphi(\tau, \sigma)=\sum_n \alpha_n(\tau) e^{i n \sigma}$ and extracting the value of $\sinh \rho $  from (\ref{ads3solutionb}) , the perturbation equation becomes
\begin{equation}
\ddot{\alpha}_n(\tau)+\left((n^2+2 q m \mathcal{E})-\frac{2\mathcal{E}^2}{a^2 sd^2(b \tau,c)}+2m^2a^2 sd^2(b \tau,c) \right)\alpha_n(\tau)=0,
\end{equation}
where $$a=\sqrt{\frac{-R_+ R_-}{R_+-R_-}}, ~~b=m (1-q^2) \sqrt{R_+-R_-}   ~~\text{and} ~~
c= \frac{R_+}{R_+-R_-}.$$
In short string limit $\mathcal{E} << m $ and for small value of q
$$a\approx \frac{\mathcal{E}}{m-q \mathcal{E}},~~~b\approx m-q \mathcal{E},~~~c\approx 0.$$ Therefore equation (\ref{ads3solutionb}) becomes
\begin{equation}
\sinh \rho \approx \frac{\mathcal{E}}{m-q\mathcal{E}} \sin (m-q\mathcal{E}) \tau.
\end{equation}
Taking the above approximation, the perturbation equation for short string  turns out to be
\begin{equation}
\ddot{\alpha}_n(\tau)+\left((n^2+2qm\mathcal{E})-\frac{2(m-q \mathcal{E})^2}{\sin^2 (m-q\mathcal{E}) \tau} \right)\alpha_n(\tau)=0,
\end{equation} 
The general solution of this second order differential equation can be written as the superposition of two linearly independent solutions. 
\begin{equation}
\alpha_n=C_1  ~P_1(\tau)+C_2 ~ P_2(\tau)
\end{equation}
where 
\begin{equation}
P_1(\tau)= \varepsilon_0 \sin^2 (m-q\mathcal{E}) \tau~~ {}_2F_1\left(1+\frac{\sqrt{n^2+2qm\mathcal{E}}}{2(m-q\mathcal{E})},1-\frac{\sqrt{n^2+2qm\mathcal{E}}}{2 (m-q\mathcal{E})},\frac{5}{2};sin^2 (m-q\mathcal{E}) \tau\right) 
\end{equation}
and
\begin{equation}
P_2(\tau)= \frac{\varepsilon_0}{\sin (m-q\mathcal{E}) \tau}~~{}_2F_1\left(\frac{\sqrt{n^2+2qm\mathcal{E}}}{2(m-q\mathcal{E})}-\frac{1}{2},-\frac{\sqrt{n^2+2qm\mathcal{E}}}{2 (m-q\mathcal{E})}-\frac{1}{2},-\frac{1}{2};sin^2(m-q\mathcal{E}) \tau\right).
\end{equation} 
The scalar function $\varphi(\tau, \sigma)$ in the perturbation equation takes the form
\begin{equation}
\varphi(\tau, \sigma)=\varepsilon_0 \sum_n \left(C_1  P_1(\tau)+C_2  P_2(\tau)\right)e^{i n \sigma}
\end{equation}
Here we can also find as similar to the previous sections that in the solution to the perturbation equation one part is finite and oscillatory while other one is diverging making the solution unstable. Finally  the physical perturbations to the background coordinates are given by

 \begin{eqnarray}
 \delta t&=& N^t \varphi(\tau, \sigma)  \nonumber \\
  &=&\varepsilon_0   \sum_n \sin (m-q \varepsilon)\tau ~  \cos  n\sigma~~\sqrt{\left(1-\frac{q\varepsilon}{m}\right)^2-\left(1-\frac{2q}{m}\right) \sin^2(m-q\varepsilon)} \nonumber \\
   &\times&{}_2F_1\left(1+\frac{\sqrt{n^2+2qm\mathcal{E}}}{2(m-q\mathcal{E})},1-\frac{\sqrt{n^2+2qm\mathcal{E}}}{2 (m-q\mathcal{E})},\frac{5}{2};sin^2 (m-q\mathcal{E}) \tau\right), \nonumber \\ 
 \delta \rho &=& N^\rho \varphi(\tau, \sigma) \nonumber \\ &=&\varepsilon_0 \sum_n  \sin (m-q\mathcal{E})\tau ~ \cos n\sigma \left[\left(1-\frac{q\varepsilon}{m}\right)-\frac{q \varepsilon}{m\left(1-\frac{q\varepsilon}{m}\right)} \sin^2(m-q\mathcal{E})\tau\right]\nonumber \\
 &\times&{}_2F_1\left(1+\frac{\sqrt{n^2+2qm\mathcal{E}}}{2(m-q\mathcal{E})},1-\frac{\sqrt{n^2+2qm\mathcal{E}}}{2 (m-q\mathcal{E})},\frac{5}{2};sin^2 (m-q\mathcal{E}) \tau\right), \nonumber  \\ 
 \delta \theta &=&N^\theta \varphi(\tau, \sigma)= 0.
 \end{eqnarray}
 It can be seen that, as expected when $q=0$, these results reduce to the results of the previous section.
 Some comments regarding the perturbation equations and their solutions are in order. The $B$-field influences the solution through its functional form as well as the parameters present there. Therefore, the perturbation equations are different as the classical field equation is different. It is easy to check that the
 physical perturbations of the pulsating strings in AdS background with flux is more stable compared to the case when there is no 
 flux.

\section{Concluding remarks}
In this paper, first we discuss the general formalism for the construction of perturbation equation using Polyakov action and in the presence of NS-NS flux. The geometric covariant quantities like normal fundamental form and extrinsic curvature tensor have been introduced to write the 
perturbation equations. By using pulsating string ansatz, we obtain the solution of the equations of motion and constraints in different subspaces of $AdS_3 \times S^3$ background.  In our analysis, we find the resulting perturbation equations in form of hypergeometric differential  equations.
The solutions to the perturbation equations  have been found to be linear combination of oscillatory and diverging part. We find the physical perturbation by taking only the oscillatory part of the solution.\par 
We wish to mention that the second order fluctuation of pulsating strings in $AdS_5\times S^5$ was discussed in \cite{Beccaria:2010zn} by
single gap Lam\'{e} form. In the present paper, we generalize the
results in the presence of NS-NS flux. Further, we write down
explicit solutions to the perturbation equations by reducing them to problems in the exactly solvable models. The covariant
formalism essentially presents the quadratic action of the string
in bosonic sector in terms of scalar quantities and covariants of
the geometry and it is an elegant method of studying the
perturbations.
This work is believed to be helpful in finding the first order correction to the energy which correspond to the anomalous dimension of the gauge theory operators in strong coupling regime. Further extension of perturbation of fermionic part of the Green-schwarz action would be worth attempting.  We wish to comeback to this issue in future.
\section*{Acknowledgement:}
We would like to thank S. P. Khastagir and Sayan Kar for some useful discussions.
\appendix
\appendixpageoff
\section{Appendix}
\subsection*{Elliptic integral and Jacobi elliptic function:}

In this appandix we collect some relevant formulas we have used in this paper. The incomplete elliptic integral of first kind is defined as

\begin{eqnarray}
\mathbb{F}(\phi, m)=\int_0^ \phi \frac{d \theta}{\sqrt{1-m sin^2 \theta}},
\end{eqnarray}
where the range of modulus m and amplitude $\phi$ are $0\leq m\leq 1$ and $0 \leq \phi \leq \frac{\pi}{2}$ respectively.
The incomplete elliptic integral of first kind becomes complete elliptic integral when $\phi=\frac{\pi}{2}$,
\begin{equation}
\mathbb{K}(m)=\int_{0}^{\frac{\pi}{2}}\frac{d \theta}{\sqrt{1-m sin^2 \theta}}
\end{equation}

Similarly the incomplete and complete integral of second kind are defined as
\begin{equation}
\mathbb{E}(m,\phi)=\int_{0}^{\phi} d\theta\sqrt{1-m \sin^2 \theta},~~~~~~~~ \mathbb{E}(m)=\int_{0}^{\frac{\pi}{2}} d \theta\sqrt{1-m \sin^2 \theta}
\end{equation}
The Jacobi elliptic functions are defined in terms of Jacobi amplitude,
\begin{equation}
\phi=am(u,m)=\mathbb{F}^{-1}(u,m)
\end{equation}
\begin{eqnarray}
sn(u,m)=\sin(am(u,m))=\sin \phi \nl \\
cn(u,m)= \cos(am(u,m))=\cos \phi \nl \\
dn(u,m)=\sqrt{1-m \sin^2 \phi}= \sqrt{1- m \sin^2(am(u,m) }.
\end{eqnarray}
Some of the useful properties and relations of Jacobi elliptic functions are \cite{Gradshteyn00}
\begin{eqnarray}
sd(u,m)= \frac{sn(u,m)}{ dn(u,m)},~~~~~~~ds(u,m)=\frac{dn(u,m)}{sn(u,m)}, \\
cn^2(u,m)+sn^2(u,m)=1, \nl \\
dn^2(u,m)+m sn^2(u,m)=1.
\end{eqnarray}
\begin{eqnarray}
sn(u,m)&=&\frac{1}{\sqrt{m}}sn(\sqrt{m}u,\frac{1}{m}),~~ cn(u,m)=dn(\sqrt{m}u,\frac{1}{m}),~~ dn(u,m)=cn(\sqrt{m}u,\frac{1}{m}), \nl \\
sn^2(u)&=&\frac{1-cn(2u)}{1+dn(2u)},~~~~~ sn(u,m)=\frac{1}{\sqrt{m} sn(u+i \mathbb{K}^\prime(m),m)}.
\end{eqnarray}
The derivatives of Jacobi elliptic functions are:
\begin{eqnarray}
\frac{d}{du}sn(u,m)&=&cn(u,m)~dn(u,m), \nonumber \\
\frac{d}{du}cn(u,m)&=&-sn(u,m)~dn(u,m), \nonumber \\
\frac{d}{du}dn(u,m)&=&-m sn(u,m)~cn(u,m).
\end{eqnarray}

\subsection*{Heun's Equation:}
Heun's equation is a second-order linear ordinary differential equation (ODE) of the form
\begin{equation}
\frac{\partial^2 u}{\partial s ^2}+\left(\frac{\gamma}{s}+\frac{\delta}{s-1}+\frac{\epsilon}{s-a}\right)\frac{\partial u}{\partial s}+\frac{\alpha \beta s-q}{s(s-1)(s-a)}u=0
\end{equation}
with the condition $\alpha +\beta +1=\gamma+\delta+\epsilon$.  Heun's equation has four regular singular points: $0, 1, a $ and $ \infty $. 
The solution to the Heun's equation can be written in terms of Heun function $H(a,q;\alpha,\beta,\gamma,\delta;s)$.

\subsection*{Lam\'{e} Equation:}

Lam\'{e} equation is a second-order ordinary differential equation comes in the method of separation of variables to the Laplace equation in ellipsoidal coordinates. Lam\'{e} equation can be expressed in terms of Jacobi elliptic functions as \cite{wang1989special},
\begin{equation}\label{lame}
\frac{d^2 \Lambda}{d \alpha^2}- \left(n(n+1) k^2 sn^2 \alpha +E\right)\Lambda=0.
\end{equation}

For different values of n the solutions and the corresponding eigenvalues are given as follows,
\begin{eqnarray}
\text{for} ~~ n &=&0: \Lambda=1,~~ E=0 \nl \\
n&=&1: \Lambda^{-1}= sn(\alpha),~~ \Lambda^0=\sqrt{sn^2(\alpha)-1},~~ \Lambda^1= \sqrt{sn^2(\alpha)-\frac{1}{k^2}}, \nonumber \\
&~&~~~~E^{-1}= -k^{2}-1, ~~E^ 0=-1, ~~ E^1=-k^2.
\end{eqnarray}


\providecommand{\href}[2]{#2}\begingroup\raggedright\endgroup

\end{document}